\documentclass[fleqn,12pt,twoside]{article}
\usepackage{espcrc1,epsfig}
\usepackage[figuresright]{rotating}
\usepackage{wrapfig}

\usepackage{graphicx}
\usepackage[figuresright]{rotating}

\newcommand{\beq}{\begin{equation}}
\newcommand{\eeq}{\end{equation}}

\newcommand{\AmS}{{\protect\the\textfont2
  A\kern-.1667em\lower.5ex\hbox{M}\kern-.125emS}}

\title{Important experimental observables at RHIC
\footnote{Invited talk given at the XVI International Conference on Ultrarelativistic 
Nucleus-Nucleus Collisions ``Quark Matter 2002'', Nantes, France,  July 18-24, 2002.}
}

\author{D. Kharzeev\address[BNL]{Nuclear Theory Group, \\
        Physics Department,\\ 
        Brookhaven National Laboratory, \\ 
        Upton, New York 11973-5000, USA}%
        \thanks{Work supported by the U.S. Department of Energy under Contract No. DE-AC02-98CH10886.}}

\begin{document}

\maketitle

\begin{abstract}
In this talk I discuss the significance of the first RHIC  measurements for establishing the properties of 
hot and dense QCD matter and for understanding the dynamics of the theory  
at the high parton density, strong color field frontier. 
Hopes and expectations for the future are discussed as well. 
\end{abstract}

\section{Introduction}

RHIC began operation in 2000, culminating over ten years 
of development and construction 
and a much longer period of theoretical speculations about the properties 
of hot QCD matter produced in nuclear collisions in the collider regime.  
This Conference presents a good occasion to look at the first experimental results 
and to discuss their meaning and significance. It is also time to think about 
the important questions which at present remain unanswered, and the future studies necessary 
to answer them.

RHIC is a machine entirely dedicated to the study of Quantum Chromo--Dynamics -- the theory 
of strong interactions. 
Asymptotic freedom of QCD \cite{Gross:1973id}, \cite{Politzer:fx} 
ensures that the dynamics of quarks and gluons at sufficiently high 
density can be addressed by weak coupling methods. This includes both the thermalized quark--gluon plasma 
at high temperature and the 
wave functions of the colliding nuclei described, at small Bjorken $x$,  
by parton saturation and the Color Glass Condensate \cite{GLR,MUQI,BM,MV,Mueller:2002kw}. 
What have we learned about these systems from experiment? Has the dense quark--gluon 
matter been produced at RHIC? I will return to these questions at the end of the talk; now, let us 
discuss what the experiment is telling us.

\section{Looking at the first RHIC data}

\subsection{Global observables}

Global observables are the most general characteristics of the collision, including particle 
multiplicity, its dependence on the centrality of the collision and on rapidity, and 
azimuthal distribution of the produced particles. The centrality is determined 
by the impact parameter in the collision; since this is not a quantity which can be measured 
directly, centrality is usually determined in terms of a certain cut in the multiplicity 
distribution; e.g., a $0-10 \%$ centrality cut means that out of a given sample, $10 \%$ of the events 
which have the highest multiplicity have been selected. 

\begin{wrapfigure}{r}{0.5 \textwidth}
\begin{center}
\vspace{-1.3cm}
\includegraphics[width=0.5 \textwidth]{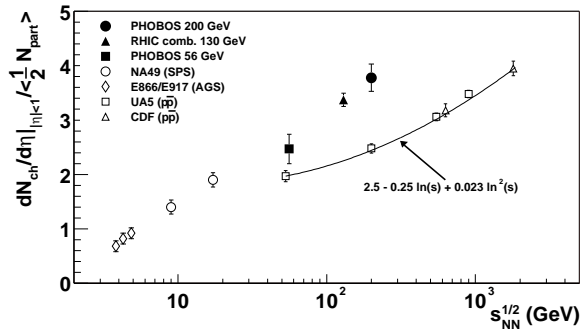}
\vspace{-0.9cm}
\end{center}
\caption{\footnotesize Centrality dependence of the charged particle multiplicity near  mid-rapidity in Au + Au 
collisions at $\sqrt{s} = 20, 130, $and $200$ GeV, from \cite{Baker2002}.}
\vspace{-0.5cm}
\label{fig:phobmult}
\end{wrapfigure}

It is convenient to characterize centrality in terms of the number of ``participants'' -- 
the nucleons which underwent an inelastic interaction in a given collision. Glauber theory 
\cite{Glauber} can be 
used to correlate a certain centrality cut with an average number of participants (for an explicit set 
of formulae for nuclear collisions, see e.g. \cite{Kharzeev:1996yx}, \cite{Kharzeev:2000ph}).
This procedure can be independently verified by measuring the energy carried forward by spectator 
neutrons; for that purpose all RHIC experiments are equipped by Zero Degree Calorimeters.  
   
\subsubsection{Multiplicity}

\begin{wrapfigure}{r}{0.5 \textwidth}
\begin{center}
\vspace{-1.3cm}
\includegraphics[width=0.5 \textwidth]{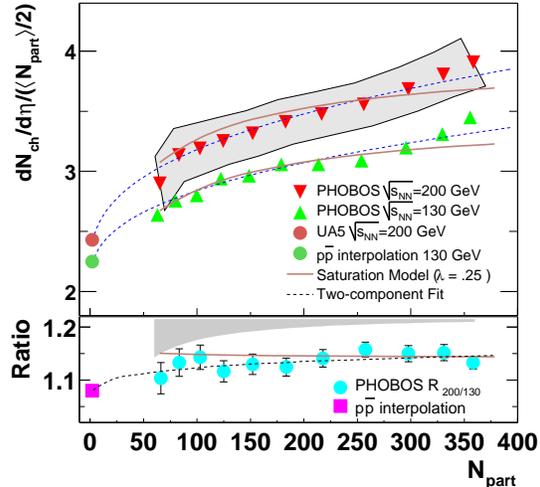}
\vspace{-0.9cm}
\end{center}
\caption{\footnotesize Centrality dependence of the charged particle multiplicity near  mid-rapidity in Au + Au 
collisions at $\sqrt{s} = 130$ GeV and $200$ GeV, from \cite{Back:2002uc}.}
\vspace{-0.5cm}
\label{fig:phobcolor}
\end{wrapfigure}

Multiplicity in heavy ion collision tells us which 
fraction of the collision energy is inelastically transferred to secondary particles. 

Theoretical expectations for hadron multiplicities at RHIC varied by factor of five 
(see, e.g., \cite{Bass:1999zq}), and 
the experimental verdict was thus eagerly awaited. After the commissioning of the machine, the first 
multiplicity results did not take long to come; they are shown on Fig. \ref{fig:phobmult} in 
comparison to the multiplicity previously measured in  $pp$ and $\bar{p}p$ collisions.

The measured multiplicity appeared much smaller than most theoretical predictions. 
Is this disappointing? To answer this question, let us recall that, by the very definition, 
an incoherent superposition of {\it independent} nucleon--nucleon collisions 
yields multiplicity equal to the number of collisions times the multiplicity in $NN$ collision. 
This trivial statement holds also in the presence of elastic rescatterings. Indeed,      
according to so-called AGK cutting rules \cite{agk} of multiple scattering theory, the
nuclear cross section is given by
\beq
E\frac{d^3\sigma^a_{AB}}{d^3p}=T_{AB}(\vec b)E\frac{d^3\sigma^a_{NN}}{d^3p}, 
\eeq
where the nuclear overlap function is 
\beq
T_{AB}(\vec b)=\int d^2s T_A(\vec s)T_B(\vec b-\vec s),
\eeq
and the nuclear thickness function $T_A(\vec b)=\int_{-\infty}^\infty dz \rho_A(\vec b,z)$ 
is the integral over the
nuclear density.
Integration over impact parameter $b$ yields
\beq
E\frac{d^3\sigma^a_{AB}}{d^3p}=AB\;E\frac{d^3\sigma^a_{NN}}{d^3p},
\eeq
and correspondingly the particle multiplicity would scale as 
\beq\label{eq:nch}
\frac{dn}{d\eta}=AB\;\frac{1}{\sigma^{in}_{AB}}
\frac{d\sigma_{NN}}{d\eta} \ \sim A^{2/3} B^{2/3} \ \frac{dn_{NN}}{d\eta}. 
\eeq

Using the numbers of collisions ($\simeq 1050$) and participants ($\simeq 340$) in central 
($0-6 \%$ centrality cut) Au Au collisions from Glauber 
 model calculations \cite{Kharzeev:2000ph}, we would thus conclude that 
$Au Au$ multiplicity per participant pair should exceed $NN$ multiplicity by factor of $3$. 
Instead, the data at the highest RHIC energy of $\sqrt{s} = 200$ GeV show the difference of 
only about $50 \%$. Given that any inelastic rescatterings in the final state can only 
increase the multiplicity, we therefore have an experimental {\it proof} of a high degree 
of coherence in multi-particle production in nuclear collisions at RHIC energies.  
The diagrams which allow to evade the AGK theorem are Gribov's ``inelastic
shadowing'' corrections \cite{glaubergribov} which correspond to the excitation of high--mass 
intermediate states in multiple scattering; the process thus no longer can be decomposed in terms 
of elementary $NN$ interactions. In parton language, these contributions correspond to 
multi--parton coherent interactions. 

\subsubsection{Centrality dependence}

The dependence of multiplicity upon the number of participants discussed above can be established by 
selecting different centrality cuts. The result is shown in Fig.\ref{fig:phobcolor}; one can see that 
the multiplicity per participant pair increases with centrality, but not quite as fast as it would 
if the $NN$ collisions were independent. 
\begin{figure}[htb]
\vspace{-1cm}
\epsfig{file=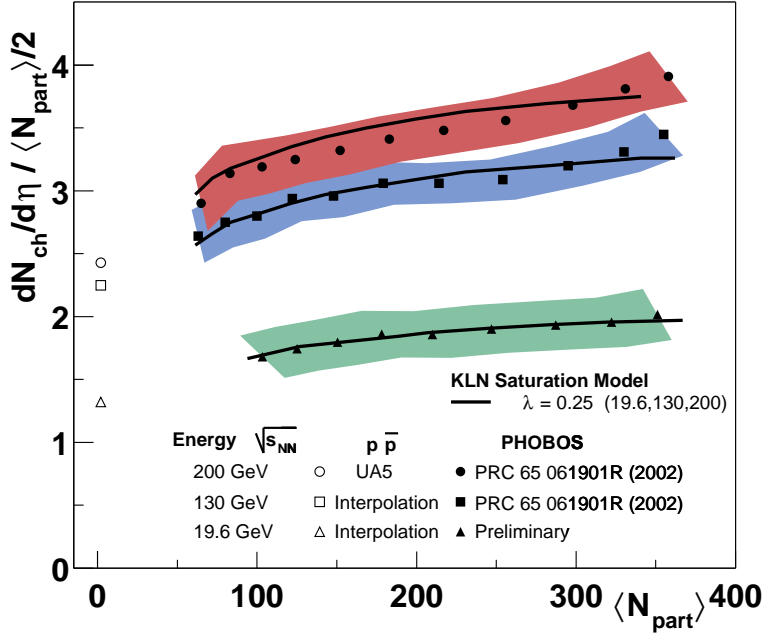,width=11cm}
\vspace{-0.5cm}
\caption{\footnotesize Centrality dependence of the charged particle multiplicity near  mid-rapidity in Au + Au 
collisions at $\sqrt{s} = 20, 130, $and $200$ GeV, from \cite{Baker2002}.}
\vspace{-0.5cm}
\label{fig:phob20}
\end{figure}

If we decompose the multiplicity measured in $NN$ collisions 
at some energy $\sqrt{s}$ into a fraction $X(s)$ coming from ``hard'' processes, 
and the remaining fraction $1-X(s)$ coming from 
``soft'' processes, and assume that in nuclear collisions ``hard'' processes are incoherent and thus 
scale with the number of collisions, whereas ``soft'' processes 
scale with the number of participants \cite{Bialas:1976ed}, we arrive at the following simple parameterization 
\cite{Kharzeev:2000ph}
\beq
{d n_{AA} \over d \eta} = \left[ (1-X(s))\ \left< N_{part} \right> + X(s) \left< N_{coll} \right>\right] \ 
{d n_{NN} \over d \eta},
\eeq
which describes the data quite well. In the framework of perturbative QCD approach, one has to assume 
that the coefficient $X(s)$ is proportional to the mini--jet production cross section, and thus 
grows with energy reflecting the growth of the parton distributions at small Bjorken $x$, 
$X(s) \sim [x G(x)]^2$, with $x \sim 1/\sqrt{s}$. 

\begin{wrapfigure}{r}{0.5 \textwidth}
\begin{center}
\vspace{-1.3cm}
\includegraphics[width=0.5 \textwidth]{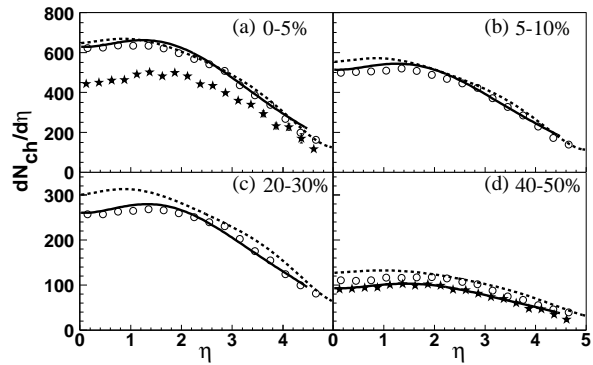}
\vspace{-0.9cm}
\end{center}
\caption{\footnotesize Pseudo-rapidity distributions of charged particles from Au + Au  collisions at 
$\sqrt{s} = 200$ GeV (open circles), from \cite{Bearden:2001qq}. Solid line is the prediction based 
on parton saturation \cite{Kharzeev:2001gp}, and dashed line is the multi-phase transport 
model calculation \cite{Zhang:1999bd}. }
\vspace{-0.5cm}
\label{fig:brahmsrap}
\end{wrapfigure}  
Therefore one expects \cite{Wang:2000bf} that 
the centrality dependence should become increasingly steep as the $\sqrt{s}$ increases (for the latest 
development, see however \cite{Li:2001xa}). This increase 
is not seen in Fig.\ref{fig:phobcolor}, which in the lower panel shows that the ratio of the  
distributions at $\sqrt{s} = 200$ GeV and  $\sqrt{s} = 130$ GeV is constant within error bars.
The almost constant ratio appears consistent with the prediction \cite{Kharzeev:2001gp},\cite{Kharzeev:2000ph} 
based on the ideas of parton saturation, where the increase of multiplicity stems from the running of the 
QCD coupling constant determining the occupation number $\sim 1/\alpha_s$ of gluons in the classical field.

A very important result presented at this Conference is the centrality dependence of charged hadron 
multiplicity at RHIC injection energy of $\sqrt{s} = 20$ GeV \cite{Baker2002}, Fig.\ref{fig:phob20}.
One can see that the shape of the centrality dependence changes very little over a large energy range, in which 
the perturbative minijet cross section grows by over an order of magnitude. The prediction of 
the saturation model \cite{Kharzeev:2001yq} is seen to agree with the data; this indicates the possibility 
that parton saturation sets in in heavy ion collisions already at moderate energies.  

\subsubsection{Rapidity distributions}

Distributions of the produced particles in the emission angle $\theta$ (with respect to the collision axis), 
or pseudo-rapidity $\eta = - \ln [\tan (\theta/2)]$ provide another important characteristic of the 
collision process. Two features of RHIC results (see Fig. \ref{fig:brahmsrap}) are especially 
noteworthy: i) the distributions do not exhibit scaling in $\eta$; ii) the deviation from $NN$ results is 
maximal in the central rapidity region whereas the shapes of the $AA$ and $NN$ distributions are similar 
in the fragmentation region. 
Moreover, when corrected for the different beam rapidity $\eta \to \eta - y_{beam}$, distributions at 
different energies exhibit approximate scaling in the fragmentation region (``limiting fragmentation''). 
\begin{wrapfigure}{r}{0.5 \textwidth}
\begin{center}
\vspace{-1.3cm}
\includegraphics[width=0.5 \textwidth]{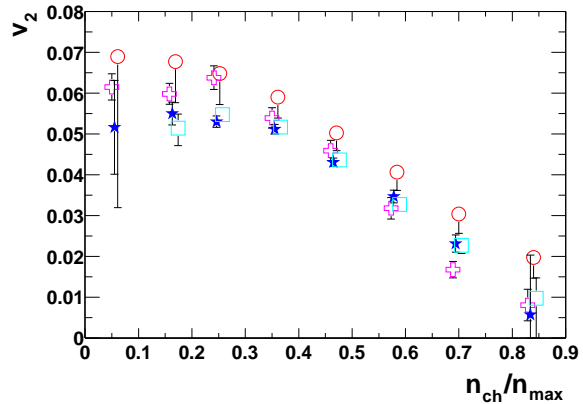}
\vspace{-0.9cm}
\end{center}
\caption{\footnotesize Azimuthal anisotropy of hadron production in Au + Au collisions at at $\sqrt{s} = 130$ GeV; 
$v_2$ is the weight of the second harmonic, $\cos 2 \varphi$, in the particle distribution in the 
azimuthal angle $\varphi$; from \cite{Adler:2002pu}. }
\vspace{-0.5cm}
\label{fig:v2star}
\end{wrapfigure}

It will be very interesting to measure rapidity and centrality dependence of hadron production at RHIC 
in $dA$ collisions. One of the really important questions which can be answered this way is whether  
there is a significant multiplication of the initially produced partons due to thermalization in 
$Au-Au$ collisions \cite{Mueller:2002kw}, or the number of the produced hadrons is directly determined by  
the number of partons liberated early in the collision from the wave functions of the colliding nuclei.
Very important constraints in this respect are provided by the measurements of the total transverse 
energy \cite{Bazilevsky:2002fz}.

\subsubsection{Azimuthal distributions}

Of great interest and importance are the distributions of the produced hadrons in the azimuthal angle. Indeed, 
if all of the $NN$ collisions were independent, there would be no reason to expect asymmetry in the 
distribution of the produced hadrons in the azimuthal angle (measured with respect to the reaction plane).

This is why the azimuthal asymmetry represents a sensitive test of the collective effects in nuclear 
collisions. The azimuthal angle distributions of the produced hadrons are usually expanded in harmonics in the 
following way\footnote{The absence of the terms proportional to $\sin n \varphi$ is the consequence of parity conservation; 
it would be interesting to search for their presence in the data in view of the speculative scenarios allowing for 
$P$ and $CP$ violation in hot QCD \cite{Kharzeev:1998kz}.}:
\beq
{d N \over d \varphi} = 1 + 2 v_1 \cos \varphi + 2 v_2 \cos 2 \varphi + ...
\eeq
Fig. \ref{fig:v2star} shows the extracted from RHIC data coefficient $v_2$ ($v_2 \neq 0$ in the language of the 
field corresponds to ``elliptic flow''). One can see that the asymmetry of the azimuthal distribution is 
quite sizable, and for peripheral collisions (small multiplicity $n_{ch}/n_{max}$) reaches about $35 \%$. 

This effect certainly indicates the presence of collectivity in nuclear collisions, and comes out 
quite naturally in hydrodynamical calculations which assume complete thermalization in the final state \cite{Teaney:2000cw}, 
\cite{Kolb:2000fh}. However, final state effects are not the only possible origin of the azimuthal asymmetry; 
indeed, as we have discussed above, the high degree of coherence in the {\it initial} state signaled by the 
multiplicity measurements, in the parton saturation scenario, introduces strong correlation between the 
transverse 
momentum of the parton and its transverse coordinate in the wave functions of the nuclei. When the nuclei 
collide, this effect mimics at least a part of the asymmetry usually ascribed exclusively 
to final--state interactions 
\cite{Kovchegov:2002nf}, \cite{Krasnitz:2002ng}. The magnitude of the elliptic flow which can originate from 
the coherence in the initial state is still a subject of ongoing research and debates.      

\begin{wrapfigure}{r}{0.5 \textwidth}
\begin{center}
\vspace{-1.3cm}
\includegraphics[width=0.5 \textwidth]{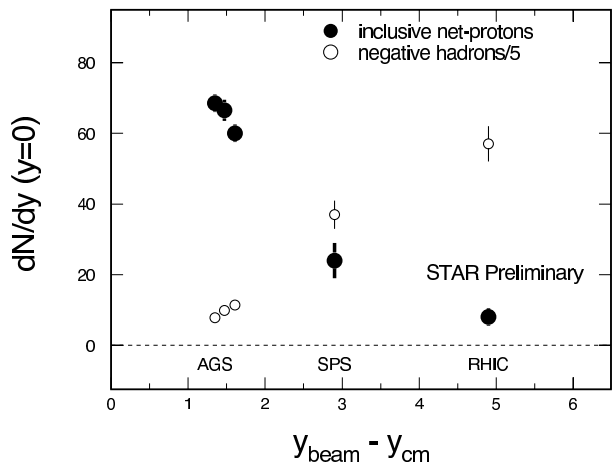}
\vspace{-0.9cm}
\end{center}
\caption{\footnotesize The yield of net protons at mid-rapidity in central collisions at different energies 
(black circles); also shown is the inclusive yield of negative hadrons (open circles); from \cite{VanBuren:2002sp}.} 
\vspace{-0.5cm}
\label{fig:starstop}
\end{wrapfigure} 
The dependence of the elliptic flow on the transverse momentum of the hadrons presents a puzzle 
 \cite{Adler:2002pu}; 
the value of $v_2$ is seen to first increase with $p_t$, and then saturate. This 
contradicts to hydrodynamics, predicting a monotonic increase of $v_2$ with $p_t$; of course 
hydrodynamics cannot be trusted at high $p_t$ anyway because the density of hard particles is 
too small to allow for a meaningful statistical description. Energy loss of the produced 
jets can contribute to the azimuthal asymmetry at high $p_t$  (see, e.g., \cite{Gyulassy:2001kr}), 
even though it is not yet clear if the magnitude of the effect can be reproduced under realistic 
assumptions about the density of the medium and the jet interaction cross section  \cite{Shuryak:2001me}. 

\subsubsection{Hadron abundances}

The measurements of yields of different hadrons at RHIC hold many surprises. Of particular interest is  
the fact that even at RHIC energies the asymmetry between baryons and antibaryons 
is still sizable, with $\bar{p}/p$ ratio about $0.65$ \cite{Adler:2001bp}; see Fig.\ref{fig:starstop}.
 This signals the diffusion of 
baryon number to quite small values of Bjorken $x \sim 10^{-2}$.  

Thermal statistical models are remarkably successful in explaining the yields of different hadrons at RHIC 
(see talks at this Conference, \cite{Ullrich:2002tq,VanBuren:2002sp}). This is certainly consistent 
with what is expected if the thermalized matter is produced. Can this success be considered a {\it proof} 
of thermalization? In my opinion, before we can conclude this we would have to understand better the origin 
of the successes of this model in explaining the particle yields in elementary $e^+e^-$ and $pp$ collisions, 
where conventional thermalization mechanisms are unlikely to operate.  

\subsection{Hard processes}
\vskip0.3cm
\subsubsection{Suppression of high $p_t$ particles}

\begin{wrapfigure}{r}{0.5 \textwidth}
\begin{center}
\vspace{-1.3cm}
\includegraphics[width=0.5 \textwidth]{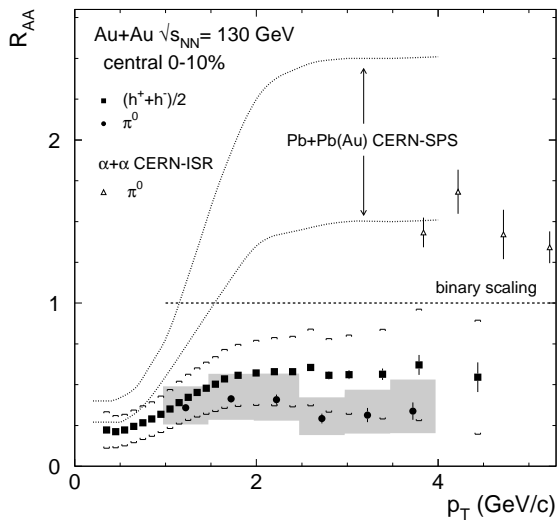}
\vspace{-0.9cm}
\end{center}
\caption{\footnotesize The ratio of transverse momentum distributions of charged hadrons and neutral pions in 
Au + Au and $pp$ collisions at $\sqrt{s} = 130$ GeV; from \cite{Adcox:2001jp}. }
\vspace{-0.5cm}
\label{fig:phenixpt}
\end{wrapfigure}
Jet energy loss was among the first signatures suggested for the 
diagnostics of the hot quark--gluon matter \cite{Bj,GW,BDMPS}. This is why the measurements 
of the high $p_t$ hadron production excited a lot of interest. Indeed, the experimental 
results are striking -- 
as can be seen in Fig.\ref{fig:phenixpt}, the yield of high $p_t$ hadrons is drastically reduced with respect 
to what is expected for incoherent production in $NN$ collisions. This behavior is very different from what 
was previously observed in $Pb-Pb$ collisions at CERN SPS and in $\alpha-\alpha$ collisions at CERN ISR 
(see  Fig.\ref{fig:phenixpt}). Does this important discovery signal jet energy loss in the quark--gluon plasma? 
The answer to this question can be given after we know the results of 
the forthcoming measurements in $p(d) A$ collisions, which would allow to separate clearly the effects 
coming from the initial state.  
 
\subsubsection{Azimuthal correlations}

\begin{figure}[htb]
\begin{center}
\vspace{-0.3cm}
\includegraphics{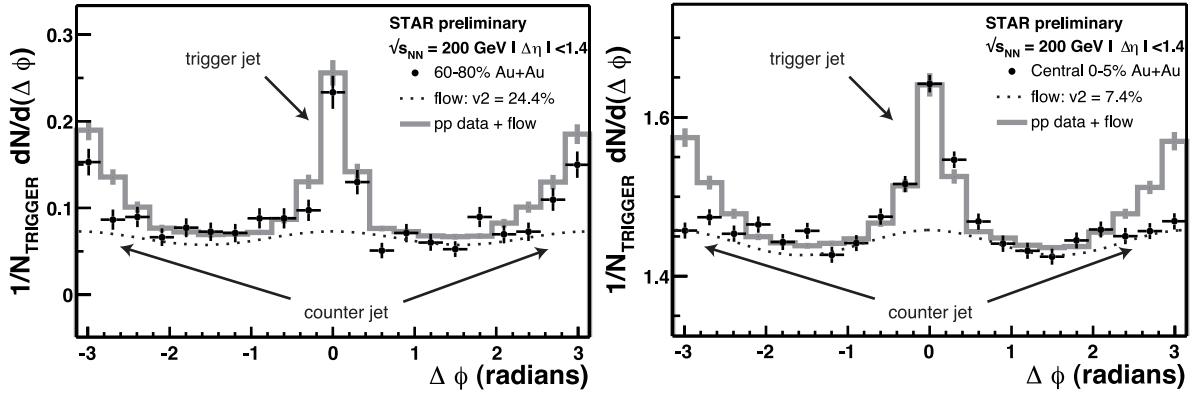}
\vspace{-0.9cm}
\end{center}
\caption{\footnotesize Azimuthal correlations of charged particles relative to a high $p_t$ trigger 
particle for peripheral (left) and central (right) collisions; from  \cite{Adler:2002ct}, \cite{Peitzmann:2002pd}.}

\vspace{-0.5cm}
\label{fig:star-jets}
\end{figure}

A very interesting recent result obtained at RHIC is the gradual disappearance of the back--to--back 
azimuthal correlations of high $p_t$ particles with centrality of the collision 
\cite{Adler:2002ct}, \cite{Kunde:2002pb}, \cite{Mioduszewski:2002wt}, 
see Fig.\ref{fig:star-jets}. This effect can be explained by the absorption of one of the  
jets in hot matter produced in central collisions. Another possible explanation stems from the onset 
of parton saturation in central collisions; this favors $2 \to 1$ parton fusion processes as the leading 
production mechanism in central collisions, and such processes produce uncorrelated mono--jets. 
Again, a definitive answer regarding the origin of this intriguing effect will be given by the measurements 
in $d-A$ collisions -- while the saturation mechanism would still operate there, the absorption mechanism 
would not.

\subsubsection{$B/\pi$ puzzle}

\begin{figure}[htb]
\begin{center}
\vspace{-0.3cm}
\includegraphics{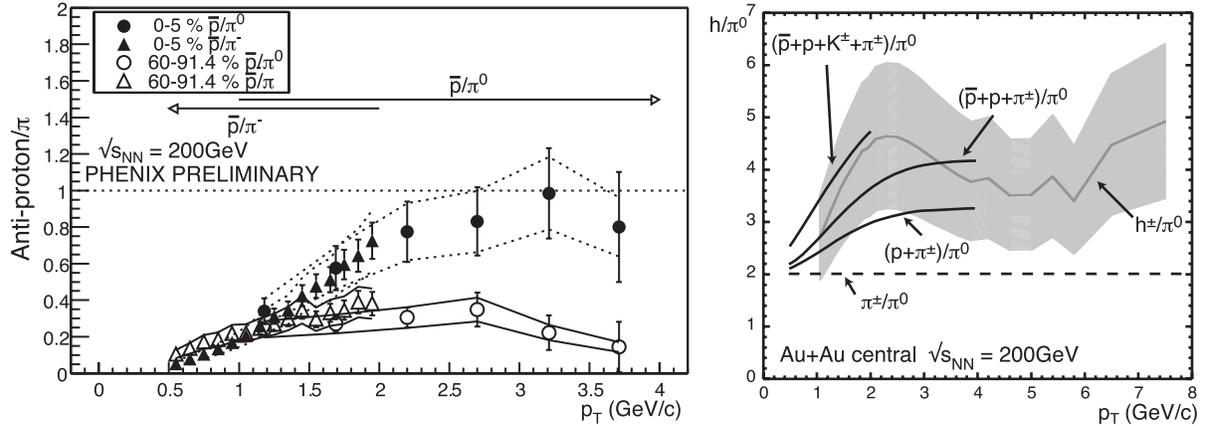}
\vspace{-0.9cm}
\end{center}
\caption{\footnotesize The ratio of antiprotons to pions as a function of transverse momentum 
for central and peripheral $Au-Au$ collisions at $200$ GeV (left), and the ratios of identified 
charged hadrons to neutral pions (right); from \cite{Sakaguchi:2002bm,Mioduszewski:2002wt,Peitzmann:2002pd}.}
\vspace{-0.5cm}
\label{fig:phenixhad}
\end{figure}

Another striking puzzle at RHIC is a rapid increase of the baryon--to--pion ratio in central 
$Au-Au$ collisions at high $p_t$ \cite{Adcox:2001mf}, see Fig.\ref{fig:phenixhad}. 
The growth of this ratio is expected in the 
hydrodynamical scenario, in which equal velocity of the expanding parton ``fluid'' implies higher 
transverse momentum for more massive particles. However, the validity of hydrodynamical description 
is dubious at high $p_t$ where the density of particles is too small.  
If we assume that minijet fragmentation is the leading production mechanism of high $p_t$ particles, then 
the growth of the $B/\pi$ ratio implies that minijet fragmentation is severely affected by the medium.

Another 
scenario \cite{Vitev:2001zn} attempts to explain both $B/\pi$ puzzle and a large value of baryon 
asymmetry $\bar{B}/B \neq 1$, and invokes the contribution of non--perturbative gluonic junctions in nuclear 
collisions \cite{Kharzeev:1996sq}. 

\subsubsection{Charm production}

\begin{wrapfigure}{r}{0.5 \textwidth}
\begin{center}
\vspace{-1.3cm}
\includegraphics[width=0.5 \textwidth]{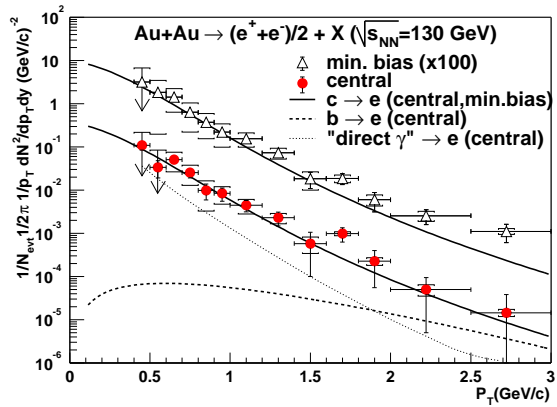}
\vspace{-0.9cm}
\end{center}
\caption{\footnotesize The background--subtracted electron spectra for minimum bias ($0-92 \%$) and 
central ($0-10 \%$) 
collisions compared with the expected contributions from open charm decays; from \cite{Adcox:2002cg}.}
\vspace{-0.5cm}
\label{fig:phencharm}
\end{wrapfigure}

The production of heavy flavors and quarkonia represents a very important and exciting part of RHIC program. 
While these studies will benefit in the future from increased luminosity and improvements in the detectors 
(allowing, in particular, to reconstruct the decay vertex of heavy hadrons), the first measurement of 
charm production cross section has been already reported \cite{Adcox:2002cg}. This has been done by deciphering the 
charm decay contribution to the single electron spectrum -- see Fig.\ref{fig:phencharm}. 

Of particular interest is the fact that while the production cross sections of light hadrons, as discussed above, 
show strong nuclear effects, the cross section of charm production, within the error bars of the measurement, scales 
with the number of $NN$ collisions. 
These results may imply much smaller, in comparison to light quarks, energy loss of heavy quarks \cite{Dokshitzer:2001zm} 
stemming from the suppression of the gluon radiation at small angles (``dead cone effect'').     

Very intriguing first results on the production of charmonium have been reported at this Conference 
\cite{Frawley:2002vz},\cite{Nagle:2002ib}. Since charmonium remains one of the best tools for the study 
of deconfinement \cite{Satz:2002ku}, it will be extremely important 
to establish the dependence of charmonium production cross section in nuclear collisions on centrality 
(for a discussion of various theoretical approaches to charmonium suppression at SPS, see e.g. 
\cite{Kharzeev:1997wa}).

\section{What have we learned so far?}
It is too early to assess the implications of RHIC results; however, it is becoming increasingly clear that 
they challenge most, if not all, of the existing theoretical dogmas. A coherent and convincing theory 
describing all of the observed phenomena directly in terms of QCD still has to be born. 
However, we can already conclude that many of the observed phenomena clearly manifest collective behavior; 
nuclear collisions at RHIC are not an incoherent superposition of nucleon--nucleon collisions.
 
The measured particle multiplicities and transverse momentum spectra allow to estimate initial 
energy density at the early moments of the collision; a typical value inferred in this way is about 
$20\ {\rm GeV/fm}^3$ (see, e.g., \cite{Kharzeev:2000ph}). The dynamics of strongly interacting matter at 
such energy density (exceeding the energy density in a nucleus by over two orders of magnitude!) 
should be described in terms of quarks and gluons, and the collective phenomena observed at RHIC thus 
directly reflect the properties of high density QCD.

\end{document}